\documentclass[aps,pra,showpacs,twocolumn,floats,epsfig,pdflatex]{revtex4}
\usepackage{epsfig}
\usepackage{epstopdf}
\usepackage{amsmath}
\usepackage{amsfonts}
\usepackage{graphicx}
\usepackage{amssymb}
\usepackage{amsbsy}
\usepackage{subfigure}

\newcommand{\ket}[1]{|{#1}\rangle}

\begin{document}

\title {Theory of phonon dynamics in an ion trap}

\author{T. Dutta,$^{(1)}$ M. Mukherjee,$^{(1,2,3)}$ K. Sengupta$^{(4)}$}
\affiliation{$^{(1)}$ Centre for Quantum Technologies, National
University Singapore, Singapore 117543, Singapore. \\ $^{(2)}$
Physics Department, National University of Singapore, 2 Science
Drive 3 Singapore 117551, Singapore. \\ $^{(3)}$ Majulab (UMI3654) International Joint Research Unit CNRS-UNS-NUS-NTU, Singapore. \\$^{(4)}$ Theoretical Physics Department, Indian Association for the Cultivation of Science,
Jadavpur, Kolkata-700032, India.}

\date{\today}

\begin{abstract}

We develop a theory to address the non-equilibrium dynamics of
phonons in a one-dimensional trapped ion system. We elaborate our
earlier results obtained in Phys. Rev. Lett. {\bf 111}, 170406
(2013) to chart out the mechanism of dynamics-induced cooling and
entanglement generation between phonons in these systems when
subjected to a linear ramp protocol inducing site-specific tuning of
on-site interactions between the phonons. We further extend these
studies to non-linear ramps and periodic drive protocols and
identify the optimal ramp protocol for minimal cooling and entanglement
generation time. We qualitatively address the effect of noise
arising out of fluctuation of the intensity of the laser used to
generate entanglement and provide a
detailed discussion of a realistic experimental setup which may
serve as a test bed for our theory.

\end{abstract}
%

\maketitle

\section{Introduction}
\label{intro}

Trapped ion systems, similar to their cold atom counterparts, are
well known emulators of specific models of strongly correlated
quantum systems such as the Ising and the Bose Hubbard models
\cite{ionising1,ionbose1}. A typical example of such a system
constitutes a linear chain of ions in a trap
\cite{ionref1,ionref2}. The physics of the phonons, which are the
motional quanta of these charged ions in the trap, can be shown to
be described, under appropriate conditions, by the Bose-Hubbard
model (BHM) \cite{ionbose1}. This leads to the realization of
one-dimensional (1D) Bose-Hubbard model in such systems. The
equilibrium properties and quantum ground states of these phonons
has already been studied theoretically in details \cite{ionbose1,ionbose2}. In
particular, it was shown that such system may be the perfect test
bed for studying the quantum phase transition between the localized
Mott and the delocalized superfluid states of the phonons
\cite{ionbose1,ionbose2, subir1}.

The key difference between emulation of such correlated model
Hamiltonians using trapped ions \cite{duan1} and ultracold atoms
\cite{bloch1,zoller1,sengupta1} are two fold. First, in contrast to
their ultracold atom counterparts, trapped ion systems provide
easier local control over the parameters of the model. For example,
the local interaction potentials between the phonons at a given site
can be experimentally tuned for individual sites of the chain.
Second, ion trap systems are usually prepared in form of 1D chains;
in contrast to their cold atom counterparts, it is not easy to
prepare coherent higher dimensional configurations of ions in traps.
The second feature makes such system an ideal choice of for studying
correlated physics in low-dimension (1D) while the first allows for
local control over systems parameters which is very difficult to
achieve in cold atoms.

A feature of the trapped ion systems which is of key interest to the
present study is that they allow one to study the non-equilibrium
dynamics of the model they emulate. This feature is somewhat similar
to ultracold atom emulator; however, the trapped ion system allow us
to dynamically change local system parameters with arbitrary
precision. Consequently, one can study the non-equilibrium dynamics
of the emulated model under local protocols. Since these ion trap
systems can be used to emulate correlated models which harbors
quantum phase transitions, they provide us with an unique opportunity
to study the effect of local non-equilibrium dynamics near a quantum
critical point \cite{ks1,ks2}. Moreover, trapped ions are ideal
system to test atomic physic at high precision \cite{DDM15} in a controlled and
clean way which makes them ideal units for realization of a quantum
computer.

In this work, we study the effect of non-equilibrium dynamics of the
Bose-Hubbard model, which describes the phonons in a linear chain of
trapped ions, under local protocols. The kinetic energy of these phonons is described by a nearest
neighbor hopping term while the interaction between them is local
(on-site). Such a Hamiltonian can be written as
\begin{eqnarray}
H = J \sum_{\langle ij\rangle} (b_{i}^{\dagger} b_j +{\rm h.c.} ) +
\sum_i U_i {\hat n}_i({\hat n}_i -1) \label{ham1}
\end{eqnarray}
where $J>0$ is the hopping amplitude, $U_i$ is the on-site
interaction between the phonons at the $i^{\rm th}$ site, $b_j$ is
the annihilation operator for the phonon at the $j^{\rm th}$ site,
and ${\hat n}_i= b_i^{\dagger} b_i$ is the boson number operator. In
what follows we discuss the non-equilibrium dynamics when $U_i$ is
ramped to $-U_i$ on either one or two of the sites in a chain. We
show that the first protocol leads to cooling of the ions to their
transverse motional ground state while the second leads to
generation of an entangled Bell state between the two sites. We note
that our proposal provides an viable alternative for cooling of a long chain of ions
which, in contrast to well-known sympathetic cooling \cite{side1}, does
not require the presence of multiple species of ions and provides a
cooling time which is independent of the electronic structure of the
involved ions. Further, our study reveals that non-equilibrium
dynamics can be utilized to generate pure many-body entangled states
which are computationally relevant. 

 It is well-known that the generation of entangled quantum
states is an essential prerequisite for quantum computation; the
present work analyzes a few "experimentally relevant" classes of possible dynamic
protocols for generating such a state. In particular, our work here
extends the analysis of Ref.\ \onlinecite{ks2} to both non-linear
and periodic drive protocols. We show that ramp protocols are more
efficient compared to periodic ones for cooling and entanglement
generation and that the optimal ramp protocol for shortest cooling
and entanglement generation time corresponds to a non-linear ramp
with exponent $\alpha \simeq 0.8$: $U_i(t) =
U_i^{(0)}(1-2(t/\tau)^{\alpha})$. Here $U_i^{(0)}$ is the initial value
of the interaction at site $i$ and $\tau$ is the ramp time. We also
show that the cooling (entanglement generation) can be reliably
achieved by single(two) site addressing, {\it i.e.}, the cooling
(entanglement) times do not drastically differ if the on-site
interaction on other sites of the chain have a near zero value.
Further, we qualitatively address the issue of the effect of the
presence of noise, arising out of fluctuation of the intensity of
the laser used to generate the on-site interaction, on the entanglement generation
and provide an estimate of the maximal effective noise temperature
beyond which such entanglement is lost. We note that single(two)
site addressing protocols are much easier to achieve experimentally
while the noise arising out of laser intensity fluctuations is an
inherent part of any ion trap system. Thus a discussion of these
issues is expected to make the theory developed here useful to
experimentalists working in ion trap systems.

The plan of the rest of the paper is as follows. First, in Sec.\
\ref{sys1}, we provide a detailed discussion of the experimental ion
trap system which can act as test bed of our theoretical idea and
discuss how such a system can lead to emulation of the Bose-Hubbard
model with Hamiltonian given by Eq.\ \ref{ham1}. This is followed by
Sec.\ \ref{comp1}, where we discuss the theoretical method used to
study the non-equilibrium dynamics of the model for several local
protocols. In Sec.\ \ref{res1}, we discuss the results obtained from
this study and identify the optimal protocols for cooling and
entanglement generation. We provide a qualitative discussion of the
effect noise on entanglement generation, summarize our results, and
conclude in Sec.\ \ref{dis1}.

\section{Ion trap system}
\label{sys1}

\begin{figure}
\rotatebox{0}{\includegraphics*[width=\linewidth]{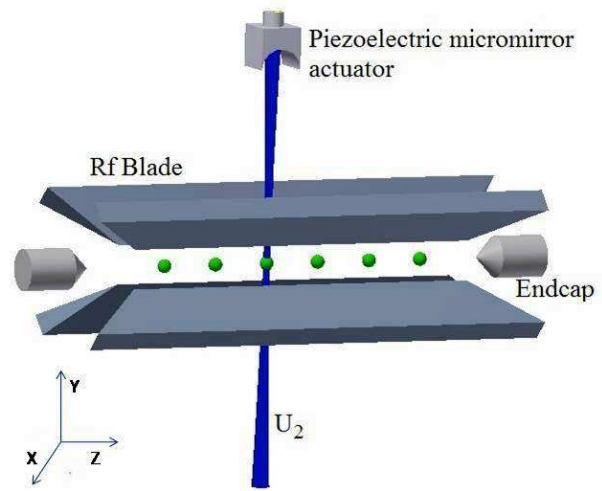}}
\caption{(Color online)Schematic of the experimental setup. In a linear chain of $6$ ions, the third ion (site no. 2) is addressed by an off-resonant laser to generate an on-site interaction $U_2$. The axis joining the two endcap electrode is considered as the $z$-axis.}
\label{figg1}
\end{figure}

In this section, we discuss the realization of an effective
Bose-Hubbard model for the transverse motional modes in a system of trapped ion linear chain starting from their vibrational dynamics. For
definiteness, we consider a string of $N$ positively charged ions
with mass $m$ confined using a linear Paul trap as shown in Fig.\
\ref{figg1} \cite{ionref1,DDM15,trap2,trap3}. The RF voltage provides a
confinement along the radial direction whereas the axial confinement
of ions is achieved by applying dc voltage at the end
electrodes. The Hamiltonian describing the vibrational motions in
such a trap is given by
\begin{equation}
H_0 = \sum_{i = 1}^N \frac{{\vec{P}}_i^2}{2 m} + V_T + V_C.
\label{vibeq1}
\end{equation}
Here $\vec{P}_i$ is the momenta of the $i^{\rm th}$ ion, $V_C$ is
the coulomb potential between any pair of the ions which can be expressed as
\begin{equation}
V_C= \frac{1}{2}\sum_{i,j=1,j\neq i}^{N} \frac{e^{2}}{\sqrt{(x_i
-x_j )^2 + (y_i -y_j )^2 + (z_i-z_j )^2}}, \label{coloumb1}
\end{equation}
where $x_i$, $ y_i$, and $z_i$ denote the coordinate of the $i^{\rm
th}$ ion in three orthogonal directions, and $V_{T}$ is the trapping
potential, which determines the equilibrium positions of the ions
and is given by
\begin{equation}
V_T = \frac{m}{2}  \sum_{i = 1}^N \left(\omega_x^2
x^2_{i}+\omega_y^2 y^2_{i}+\omega_z^2 z^2_{i} \right) .
\label{ptrap1}
\end{equation}
Here $\omega_x,\omega_y,\omega_z $ are the radial and longitudinal
trap frequencies. In what follows, we are going to work in the
regime $\omega_x,\omega_y \gg \omega_z$ so that the ions are
strongly confined in the radial direction. In our setup, the chain
direction is taken to be along $z$ as shown in the Fig.~\ref{figg1}. The oscillations of ion along
the radial directions, in this regime, is small compared to the
inter-ion distance.

Both the trapping potential and Coulomb repulsion determines the
equilibrium position of ions. In the regime described above, the
Coulomb potential is much weaker compared to the radial confinement,
and the displacements of ions in any direction can be considered to be
decoupled from the other two orthogonal directions allowing us to
discuss the vibrational modes in the radial direction only. To this
end, we expand $V_C$ as a function of radial displacement around the
equilibrium ion position and obtain the Hamiltonian
\begin{eqnarray}
H_{x} &=& \sum_{i=1}^N \Big[ \frac{P^2_{i,x}}{2 m} +
\frac{m}{2} \omega_x^2  \delta x_i^2 \Big] + \frac{1}{2}
\sum^N_{\substack{i,j = 1
\\ (i > j)}} \frac{e^2 (\delta x_i - \delta x_j)^2}{|d_{ij} ^0|^3}, \nonumber\\
\label{hmeq}
\end{eqnarray}
where $|d_{ij} ^0| = |z_i^0 - z_j^0| = d_0 |i_z-j_z|$ is the mean
distance between $i^{\rm th}$ and $j^{\rm th}$ ion along the
$z-$direction, $i_z (j_z)$ are the site number of the $i^{\rm
th}(j^{\rm th})$ site, $d_0$ is the inter-site distance between
neighboring ions, and $\delta x_i$ is the radial displacements of
the $i^{\rm th}$ ion around its radial equilibrium positions. The second
quantized form of this Hamiltonian can be obtained following
standard methods \cite{kittel1} and takes the following form ($\hbar
= 1$):
\begin{eqnarray}
H_{x} &=& \sum^{N}_{\substack{i,j = 1 \\ i > j}}
J_{ij}(b^{\dagger}_i b_j+h.c.) +
\sum^{N}_{i=1}(\omega_x+\omega_{x,i}) b^{\dagger}_i b_i \label{bhm2}
\end{eqnarray}
Here $b_i$ denotes the annihilation operator for phonons (quanta of
vibrational displacement modes) in the radial direction, and the
hopping amplitude $J_{ij}$ is induced by the Coulomb interaction.
The effective trapping frequency, $\omega_{x,i}$ depends on the
ion's positions, as well as the tunneling amplitudes $J_{ij}$ and is
given by
\begin{eqnarray}
\omega_{x,i} &=& \omega_x - \sum_{\substack{j = 1 \\ (j \neq i)}}^N
\frac{ e^2 }{2 m \omega_x |d_{ij} ^0|^3},  \quad J_{ij} = \frac{
e^2}{2 m \omega_x |d_{ij} ^{0}|^3} . \label{define1}
\end{eqnarray}
In this context, it is customary to define the dimensionless
parameter $\beta_x$ as the ratio of the Coulomb energy between
near-neighbor sites to the trapping potential energy:
\begin{equation}
\beta_x = e^2/(m \omega_x^2 d^3_0) \label{betadef}.
\end{equation}
The parameter $\beta_x$ characterizes the normal modes of the system
in radial direction; when $\beta_x
>>1$, the Coulomb repulsion dominates over the trapping potential.
In this case, the long-range nature of the Coulomb interaction
becomes important and the effect of the collective motion along the
axial direction is expected to play important role in the collective
motion of the ions and can not be ignored. However, in the limit
$\beta_x << 1$, the Coulomb interaction is effective between nearest
ions only. In this regime, which we focus on, the number
non-conserving term in the effective Bose-Hubbard Hamiltonian such
as ($b_i b_j$, $b_i^\dagger b_j^\dagger$) can be neglected since
they lead to an energy cost which is larger than the Coulomb energy.
We have used this fact to arrive at Eq.\ \ref{bhm2}. Also, since the
Coulomb interaction is effective only between the nearest
neighboring sites, one can express $J_{i,j}$ in terms of $\beta_x $
as
\begin{equation}
J_{ij}\approx J \delta_{i,j\pm 1},\quad J = \beta_x \omega_x / 2 .
\label{jdef}
\end{equation}

Note that for $\beta_x \ll 1$, the effect of on-site phonon-phonon
interaction is negligible due to the small amplitude of
anharmonicity in the trap which arises from the imperfection of trap
architecture. However, in a realistic system, this effect can be
enhanced and controlled by introducing off resonant standing wave
along  the transverse direction of the trap axis. The bosonic
Hamiltonian, in the presence of such a standing wave, has an
additional term
\begin{eqnarray}
H_{I} = F\sum\cos^2(k x_i+\frac{\pi}{2}\delta), \label{adterm}
\end{eqnarray}
where $F$ is the peak AC stark shift and $k$  is the
wavevector of the wave. This controllable anharmonicity brings the
phonons closer by making unequal the otherwise equally spaced energy
levels of radial trap potential. To quantify this effect, one may
define the dimensionless Lamb-Dicke parameter $\eta_x= k\sqrt{\hbar/2m\omega_x}<<1$; in terms of $\eta$, it is possible to write $H_I$ as
\begin{equation}
H_{I}=F\sum_{i=1}^{N}[1+\eta_x^{2}(b_{i}^{\dagger}+b_i)^{2}
+\frac{1}{3}\eta_x^{4}(b_{i}^{\dagger}+b_i)^{4}+\emptyset(\eta_x)^{6}].
\label{hidef}
\end{equation}
In the regime where $F\eta^{2}<<\omega_{x}$, the effective
interaction between the bosons is quartic and can be written as
\begin{equation}
H_{I}=(-1)^{\delta}F\eta_x^{4}/3\sum_{i=1}^{N}(b_{i}+b^{\dagger}_{i})^2\backsimeq
\sum_{i} U_{i} b_{i}^{\dagger 2}b_{i}^{2},
\end{equation}
where in arriving at the last relation we have used the fact that
the number non-conserving terms such as
$b_{i}^{2}$,$b^{\dagger^{2}}_{i}$ can be neglected since they
corresponds to much higher energy scales as discussed earlier. Also,
it can be shown that the rest of the phonon conserving terms will
induce a small shift in the radial potential which amounts to a
redefinition of the the overall potential. The above discussion
leads to the final form of the Bose-Hubbard Hamiltonian given by
Eq.\ \ref{ham1} with $J>0$ and $U_i = 2 (-1)^{\delta} F \eta_{x}^4$.
The on-site interaction strongly depends on the strength of the
standing wave and relative position($\delta=0$ or $1$) of ion on the
standing wave. Note that it is possible to have $U_i$ to be
repulsive or attractive which is one of the key facts used in
the present work. Finally, it is useful to note that the sign of the
hopping term in the present Bose-Hubbard model is positive in
contrast to its ultracold atom counterpart where such a term comes
with a negative sign.

\section{Method for studying non-equilibrium dynamics}
\label{comp1}

In this section, we outline our method for studying non-equilibrium
dynamics of the 1D Bose-Hubbard model (Eq.\ \ref{ham1}) derived
earlier. We begin by specifying the protocols for the dynamics that
we study. We start with the system in the ground state of the
Hamiltonian given by Eq.\ \ref{ham1} for a fixed $J$ and $U_i$. We
consider a chain of $N$ ions; the choice of $N$ depends on
experimental setting which is to be discussed in Sec.\ \ref{dis1}.

The dynamics is induced in the model by changing the local
interaction parameter $U_i^{(0)}$ to $U_i^{(1)}=-U_i^{(0)}$ on a one or more selected
site(s). For cooling, we change $U_i$ at any one of the $N$ sites,
while for generating entanglement this change is made for local
interactions on two of the sites of the chain. The protocol used to
make such a change are either linear or non-linear ramp or periodic
and are given by
\begin{eqnarray}
U_i(t) &=& U_i^{(0)}[1-2(t/\tau)^{\alpha}], \quad  0\le t\le \tau,  \quad
({\rm
ramp}) \nonumber\\
&=& U_i^{(0)} \cos(\pi t/\tau),\quad 0\le t\le \tau,  \quad ({\rm
periodic}), \label{protocol}
\end{eqnarray}
where the exponent $\alpha$ specifies the power-law which is followed
by the ramp and $\alpha=1$ indicates linear ramp protocol studied in
Ref.\ \onlinecite{ks2}. We note here that $\tau \to 0$ indicates the
sudden quench limit; such dynamics for the model was studied in
Ref.\ \onlinecite {ks1}. In case of periodic ramp, we would be using $T$ and $\omega_0$ to denote the period and frequency of the ramp which is related to $\tau$ as $\tau=T/2=\pi/\omega_0$.

To study the dynamics, we begin at $t=0$ assuming that the system is
in the ground state of the Bose-Hubbard model given by Eq.\
\ref{ham1} with $U_i^{(0)}=U^{(0)} > 0$ at all sites and a fixed $J/U^{(0)}$ at
each site. Since we work in a regime where changing the number of
phonons is a high-energy process, we shall study the dynamics at
fixed number $N_0$ total phonons (bosons). Thus the Hilbert space of
the bosons can be truncated to keep those states for which $n_i \le
N_0$ for any site $i$. We use exact diagonalization method for the
finite-size system within this truncated Hilbert space to obtain the
energy eigenstates $|\alpha\rangle$ and eigenvalues $E_{\alpha}$ for
$H(t=\tau)$. This amounts to a choice of basis; in terms of these
eigenstates, one can express the initial ground state (at $t=0$)
\begin{eqnarray}
|\psi_G\rangle = \sum_{\alpha} c_{\alpha}^0 |\alpha\rangle,
\label{gndcoeff}
\end{eqnarray}
where the coefficients $c_{\alpha}^0$ denote the overlap of the
initial ground state of the system, also obtained using exact
diagonalization in the same truncated Hilbert space with $H=H(t=0)$,
with $|\alpha\rangle$.

The time-dependent Schr{\"o}dinger equation for the system
wavefunction
\begin{eqnarray}
|\psi(t) \rangle = \sum_{\alpha} c_{\alpha}(t) |\alpha \rangle
\label{wav1}
\end{eqnarray}
governing the dynamics of the system now reduces to equations for
time evolution of $c_{\alpha}(t)$:
\begin{eqnarray}
i \hbar \partial_t \sum_{\alpha} c_{\alpha}(t) |\alpha\rangle = H(t)
\sum_{\alpha} c_{\alpha}(t) |\alpha\rangle \label{sch1}
\end{eqnarray}
with the boundary condition $c_{\alpha}(0)=c_{\alpha}^0$. To solve
these equations, it is convenient to rewrite
\begin{eqnarray}
H(t) &=& H(\tau) + \Delta H(t), \nonumber\\
\Delta H(t) &=& \sum_i (U_i(t)-U_i(\tau)) {\hat n}_i ( {\hat n}_i
-1). \label{sch2}
\end{eqnarray}
With this choice, one obtains the final set of equations for
$c_{\alpha}(t)$ to be
\begin{eqnarray}
(i \hbar \partial_t - E_{\alpha}) c_{\alpha}(t) &=& \sum_{\beta}
\Lambda_{\alpha \beta}(t) c_{\beta} (t) \nonumber\\
\Lambda_{\alpha \beta}(t) &=& \langle \beta | \Delta H(t) |
\alpha\rangle. \label{sch3}
\end{eqnarray}
The set of coupled equations for $c_{\alpha}(t)$ is then solved
numerically leading to an exact numerical solution for the
time-dependent boson wavefunction $|\psi(t)\rangle$. We note here
that it is clear from the structure of Eq.\ \ref{sch3} that
$c_{\alpha}(t)$ can be written as
\begin{eqnarray}
c_{\alpha}(t) &=& {\tilde c}_{\alpha}(t) e^{-i  E_{\alpha} t/\hbar}
\nonumber\\
i \hbar \partial_t {\tilde c}_{\alpha}(t) &=& \sum_{\beta} e^{i
(E_{\beta}-E_{\alpha})t/\hbar} \Lambda_{\alpha \beta}(t) {\tilde
c}_{\beta} (t) \label{ch4}
\end{eqnarray}

Having obtained the wavefunction $|\psi(t)\rangle$, one can compute
the expectation value and equal-time correlation functions
corresponding to any boson operator as a function of time. This is
given, for a local boson operator ${\hat O_j}$, as
\begin{eqnarray}
O_j(t) &=& \langle \psi(t)|{\hat O}_j|\psi(t) \rangle
\nonumber\\
&=& \sum_{\alpha \beta} {\tilde c}_{\beta}^{\ast}(t) {\tilde
c}_{\alpha}(t) e^{i (E_{\beta}-E_{\alpha})t/\hbar} \langle \beta|
{\hat O}_j|\alpha
\rangle \nonumber\\
C_{jk}^{n m;\, O}(t) &=& \langle \psi(t)|{\hat O}_j^n {\hat O}_k^m
|\psi(t) \rangle \label{corr1} \\
&=& \sum_{\alpha \beta} {\tilde
c}_{\beta}^{\ast}(t) {\tilde c}_{\alpha}(t) e^{i
(E_{\beta}-E_{\alpha})t/\hbar} \langle \beta| {\hat O}_j^n {\hat
O}_k^m|\alpha \rangle, \nonumber\
\end{eqnarray}
where $n,m$ are arbitrary integers and $j,k$ denote site indices.
Although our method is powerful enough to allow us to compute the
expectation and correlation corresponding to any bosonic operator
${\hat O}$, in what follows, we shall be mainly interested in
obtaining these quantities pertaining the boson annihilation
operators $b_j$ and the number operators ${\hat n_j} =
b_j^{\dagger}b_j$.

Before ending this section, we briefly comment about the single/two-site
addressing protocol which we shall use in Sec.\ \ref{res1}. For such
protocol $U_i^{(0)}=U^{(0)}$ at $t=0$ only on the site(s) where they are
dynamically changed during the protocol. For the rest of the sites
$U_i=0$ at all times. This protocol has the advantage of
experimental simplicity, and as we shall, see leads to faster
cooling and entanglement generation. We note here that the formalism
developed here can be applied for this protocol without any major
modification.

\section{Results}
\label{res1}

In this section, we present a detailed account of the results
obtained by numerical solution of Eqs.\ \ref{ch4} followed by
evaluation of appropriate expectation value or correlation function
by computing Eq.\ \ref{corr1}. In Sec.\ \ref{cool1}, we discuss
these results in the context of cooling. This is followed by Sec.\
\ref{entang1} where we discuss entanglement generation.

\subsection{Cooling of ions}
\label{cool1}

\begin{figure}[bbb]
\rotatebox{0}{\includegraphics*[width=\linewidth]{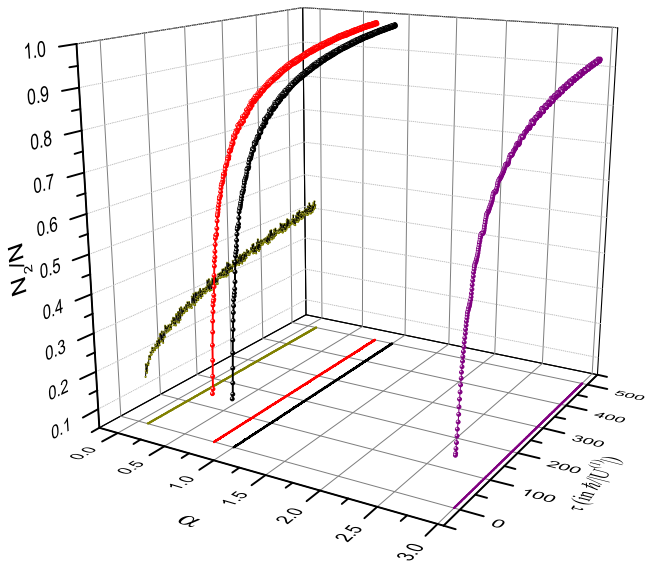}}
\caption{(Color online) Expectation values of $N_{2}/N$ as a function of the ramp time
$\tau$ and exponent $\alpha$. Linear ramp is represented by $\alpha =1$ while the other three are for exponents $0.1, 0.8$ and $3$ respectively. For all plots the initial $J/U_i^{(0)}=0.2$. The maximal phonon transfer occurs when $\alpha=0.8$(red).}\label{fig2}
\end{figure}
\begin{figure}[hhh]
\rotatebox{0}{\includegraphics*[width=\linewidth]{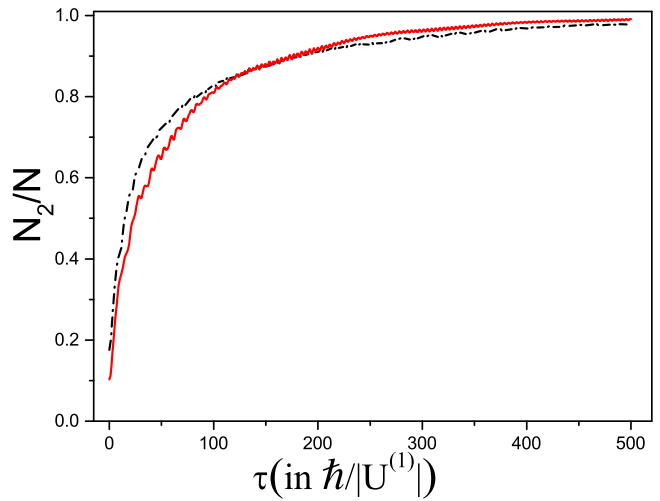}}
\caption{(Color online)  Comparative plot of cooling in case of all site except site no. $2$ applied with a repulsive on-site interaction ($U_i>0$) and no on-site interaction ($U_i=0$). The dashed-dot curve (black) is same as in Fig.~\ref{fig2} for the optimal exponent $\alpha=0.8$ while the solid-line curve is for no on-site interaction except at site number $2$. The cooling time remains similar while the final state is cooler in case of no interation as comapred to interaction applied to all sites.} \label{fig3}
\end{figure}

To achieve cooling of the ions, we follow the protocol developed in
Ref.\ \onlinecite{ks2} and change the on-site interaction on one of
the lattice sites in the chain from $U_i^{(0)} \equiv U^{(0)}$ to
$U_i^{(1)} \equiv -U^{(0)}$ with a fixed ramp rate $\tau^{-1}$. In
Ref.\ \onlinecite{ks2}, we have studied the linear ramp protocol.
Here we concentrate on non-linear ramp protocol characterized by an
exponent $\alpha$ (Eq.\ \ref{protocol}) with the aim to optimizing
the cooling time. For the sake of definiteness, we shall choose a
chain of $L=8$ site and choose to ramp the interaction on the second
site of the chain ($i=2$). We note that this choice is made for the
sake of definiteness and do not lead to loss of generality. In what
follows, we study the time evolution of the phonon population on the
second site: $N_2/N = \langle b_2^{\dagger} b_2 \rangle/N$. In the
rest of this section, we shall set the total phonon number $N=4$.

The results of such a study are shown in Fig.\ \ref{fig2}.. Fig.\
\ref{fig5}. In Fig.\ \ref{fig2}, we plot the phonon number $N_2$ at
the end of the ramp ($t=\tau$) as a function of the ramp time $\tau$
for several $\alpha$ and $J/U^{(0)}=0.2$. We find that, in accordance
with our expectation, $N_2$ approaches $N$ for slower ramps
indicating the migration of the phonons to the second site of the
chain leading to cooling of the other sites. By studying the
evolution of the phonon number on the second site, we find that
$N_2$ approaches $N$ in the shortest possible time for $\alpha
\simeq 0.8$. Thus our study reveals that a non-linear ramp with
$\alpha=0.8$ is the optimal protocol for cooling. We find that with
$|U^{(1)}|=2.8$kHz on all sites, the linear ramp achieves $N_2/N =0.9
(0.97)$ for $t=12 (26)$ms; in contrast a ramp with $\alpha=0.8$ with
same $|U^{(1)}|$ leads to $N_2/N =0.9 (0.97)$ for $t=11 (24)$ms.

Next, to study other aspects of the cooling dynamics, we resort to
single site addressing, {\it i.e}, we keep the initial interaction
parameter $U_i^{(0)} = U^{(0)} \delta_{i2}$ finite only at the site
where it is to be dynamically changed to $-U^{(0)}$; interaction is
set to zero for phonons on rest of the sites. The motivation behind
such single site addressing is that it is experimentally a lot
simpler; further as shown in Fig.\ \ref{fig3}, such single site
addressing does not lead to appreciable change in the nature of
$N_2/N$ as a function of $\tau$. In fact, the cooling time
corresponding to $N_2/N=0.97$ can be further reduced by such single
site addressing: for $\alpha=0.8$ and $|U^{(1)}|=2.8$~kHz, one reaches
$N_2/N =0.9 (0.97)$ in $t=10 (19)$ms. In the rest of this section,
we shall thus concentrate on single site addressing.

We further look for optimization of the parameter $J/U^{(0)}$ within
single site addressing. To this end, we study the behavior of
$N_2/N$ at the end of the ramp ($t=\tau$) as a function of $J/U^{(0)}$
for several representative values of $\tau$ and for $\alpha=0.8$.
Our study reveals that the maximal value of $N_2/N$ occurs at $J/U^{(0)}
\simeq 0.2$. This leads us to identify $\tau \ge 500$, $\alpha\simeq
0.8$, and $J/U^{(0)} \simeq 0.2$ as optimal ramp parameters for
achieving cooling.

\begin{figure}[hhh]
\rotatebox{0}{\includegraphics*[width=\linewidth]{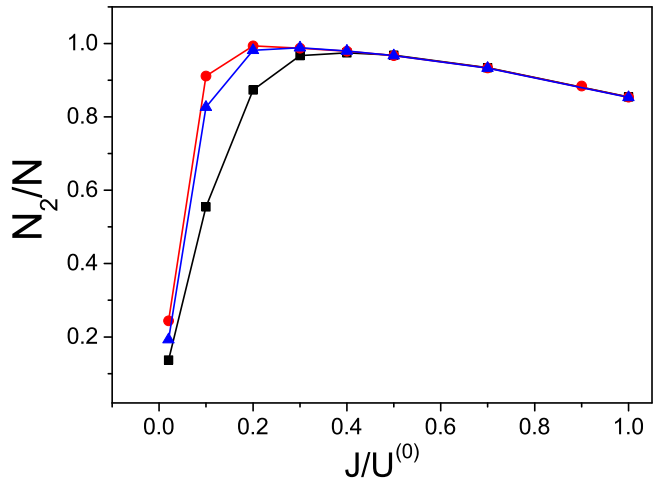}}
\caption{(Color online)Fractional photon transfer $N_2/N$ as a
function of initial BH parameter $J/U^{(0)}$ for three different
ramp time $\tau=100~\text{(square,black)},300~\text{(triangle, blue)},500~\text{(circle, red)}$ (in units of $1/|U^{(1)}|$). The ramps are
non-linear with optimized value of $\alpha=0.8$. In all these cases, the on-site interaction has been applied to a single site $2$.}\label{fig4}
\end{figure}


Finally, we compare ramp protocols to periodic protocols, described
by Eq.\ \ref{protocol},  for identifying the optimal protocol for
cooling. In what follows, we study the evolution of $N_2/N$ under
such a protocol for $J/U^{(0)}=0.2$ and with single site addressing. The
result of such a study is shown in Fig.\ \ref{fig5}. In the top left
panel of Fig.\ \ref{fig5}, we plot $N_2/N$ at the end of the drive
as a function of the drive frequency $\omega_0$. We find
that $N_2/N$ has an oscillatory behavior as a function of $\omega_0$
for small changes in $\omega_0$; however, for a larger variation of
$\omega_0$, it decreases with increasing $\omega_0$. The rest of the
panels of Fig.\ \ref{fig5} shows the time evolution of $N_2/N$ for a
wide range of $\omega_0$. From these plots, we find that one needs a
much longer cooling time if periodic protocols are used; for
example, $ N_2/N = 0.9 (0.97)$ for $t=81(88)$ms with $
\omega_0=0.0012$ in terms of $|U^{(1)}|$, $|U^{(1)}|=2.8$kHz and singe site addressing. Thus we
conclude that ramp protocols perform better than their periodic
counterparts for the purpose of cooling. The origin is this behavior
can be qualitatively understood as follows. For efficient cooling,
one needs to maximize the overlap of the final system wavefunction
with the ground state wavefunction. Such an
overlap depends on the slope of $U_i(t)$ at each instant of time.
This can be understood by noting that a near zero slope of $U_i(t)$ is
expected to reproduce the adiabatic evolution which will lead to
maximal overlap. The slope of $U_i(t)$ can be optimized more
efficiently for ramp protocols by varying the exponent $\alpha$. For
periodic protocol such as the one described by Eq.\ \ref{protocol},
the short time ($\omega_0 t \ll 1$) evolution always occur with a
near-zero slope, followed by a steeper increase at intermediate and
later times which leads to worse overlap compared to a ramp with optimal exponent.We also
note that the ramp protocol advocated here is not necessarily the
absolute optimal protocol for the problem \cite{kss}; however, it is
certainly the most efficient among the experimentally accessible
protocols discussed here.

\begin{figure}[bbb]
\vspace{0.05cm}
\rotatebox{0}{\includegraphics*[width=\linewidth]{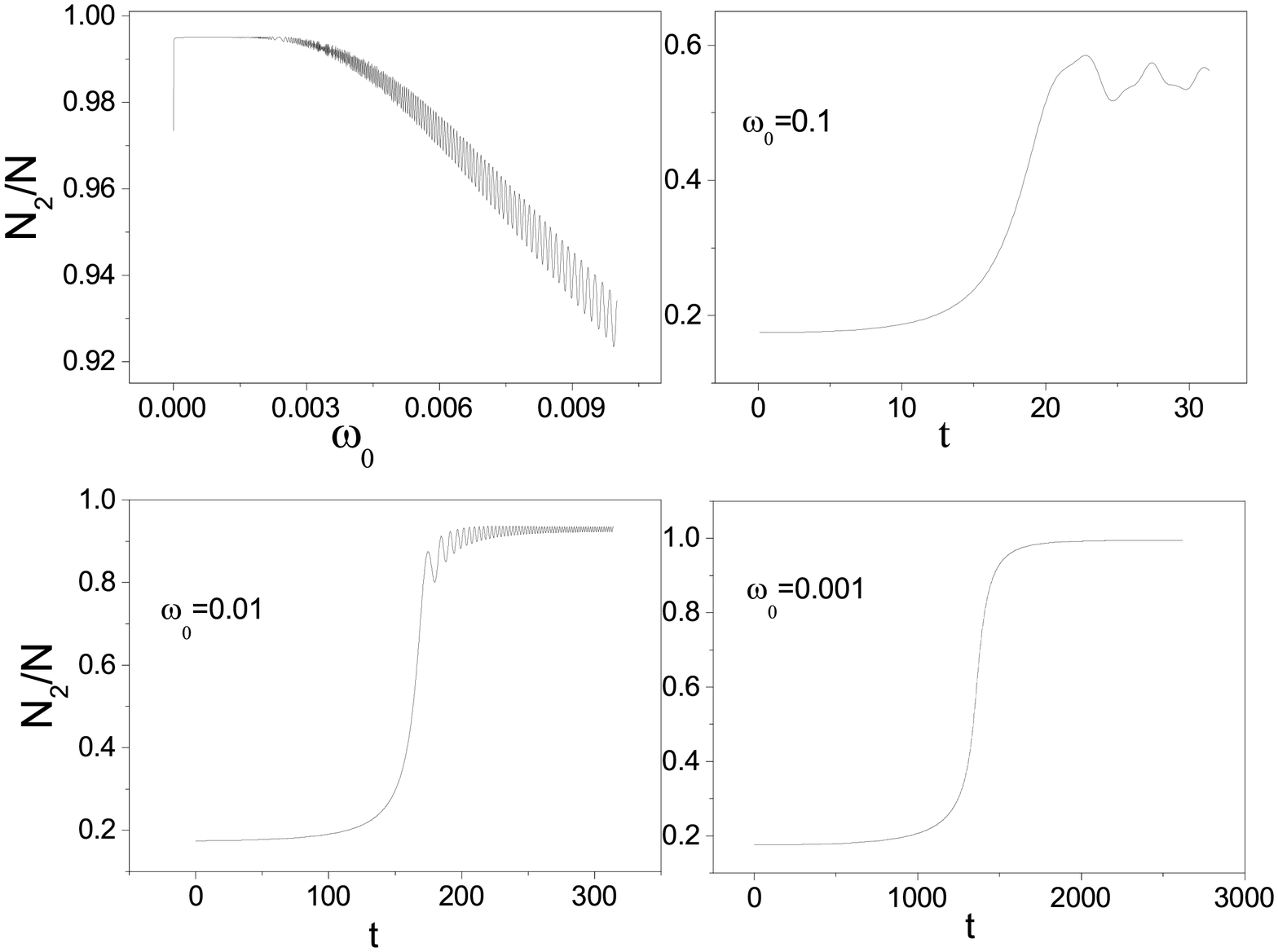}}
\caption{(Color online) Plot of $N_2/N$ for periodic ramps:
Initial $J/U^{(0)}=0.2$ and single site addressing done for all
plots. Top-left panel is $N_2/N$ vs $\omega_0$ and rest
of the figures show $N_2/N$ as a function of $t$ for
three different values of $\omega_0:$ 0.1, 0.01, and 0.001 respectively.}\label{fig5}
\end{figure}

\subsection{Entanglement generation}
\label{entang1}

\begin{figure}[bbb]
\rotatebox{0}{\includegraphics*[width=\linewidth]{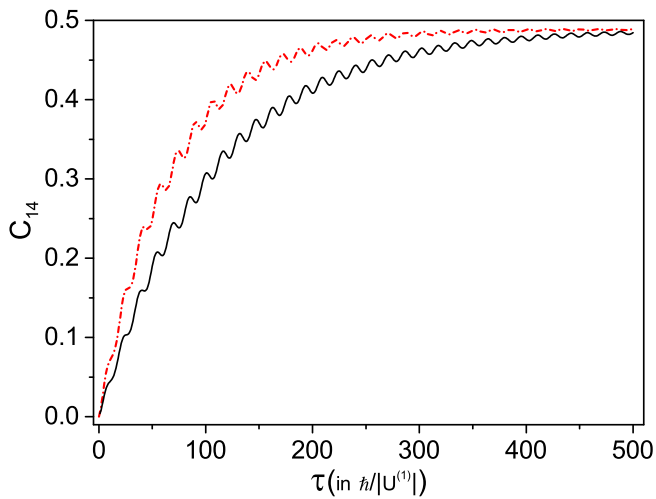}}
\caption{(Color online) Phonon-phonon correlation between sites $1$ and $4$ $C_{14}$ as a function of the ramp time
$\tau$ in units of $U^{(0)}$ for a non-linear ramp of exponent
$\alpha=0.8$. The dashed (red) curve is for $J/U^{(0)}=0.15 $ applied to
sites $1$ and $4$ only, while at the other sites $U_i=0$. The solid-line (black)
curve is for all sites initially at $J/U^{(0)}=0.15$ while ramp is
applied to only sites $1$ and $4$ while others are kept at
$J/U=0.15$. As in case of cooling (see Fig.~\ref{fig3}), application of on-site interaction to all sites during the ramp doesnot make any significant difference.} \label{fig6}
\end{figure}

In this subsection, we discuss dynamical entanglement generation
using the ramp protocol. In doing so, we choose to change the sign
of the interaction following the ramp protocol given in Eq.\
\ref{protocol} on two symmetrically placed sites on the chain. For
definiteness, and without loss of generality, in the rest of this
section, we shall choose a chain of length $L=6$ with site index
$0\le i\le 5$ and change the on-site interaction for $k=1$ and
$l=4$. We measure the cross correlation between the these two sites
which is given by
\begin{eqnarray}
C_{kl}(t) = \langle \psi(t)| (b_k^{\dagger} b_l)^N|\psi(t)\rangle/N!
\label{crreq}
\end{eqnarray}
We note that if we carried out the protocol of changing the sign of
interaction for sites $k$ and $l$ adiabatically (with infinitely
slow rate; $\tau \to \infty$), $C_{14}(t_f=\infty) = C_{14}^{\rm
max} =1/2$. Such a correlation, in the present context, pertains to
the Bell state between the sites $k$ and $l$ given by
\begin{eqnarray}
|\psi_{\rm Bell}\rangle &=&
\frac{1}{\sqrt{2}}\left(\ket{0N0000}+\ket{0000N0}\right),
\label{bells}
\end{eqnarray}
since it yields $C_{14}=C_{14}^{\rm max}$. Thus in the adiabatic
limit, $|\psi\rangle$ generated by the protocol described above is
expected to have perfect overlap with the Bell state. However, if the
ramp is done fast, as shown in Ref.\ \onlinecite{ks2}, $C_{14}(t_f)$
does not reach $C_{14}^{\rm max}$ even when $t_f \gg \tau$. Since
one of the motivations behind generation of entangled state is
performing quantum gate operations in the shortest possible time, it
is therefore useful to find optimal ramp time $\tau$ and power
$\alpha$ for which $C_{14}(t_f)$ reaches a significant fraction of
$C_{14}^{\rm max}$. This issue has been investigated for linear ramp
protocols in Ref.\ \onlinecite{ks2}; here we generalize such study
to non-linear ramp protocols.
\begin{figure}[hhh]
\rotatebox{0}{\includegraphics*[width=\linewidth]{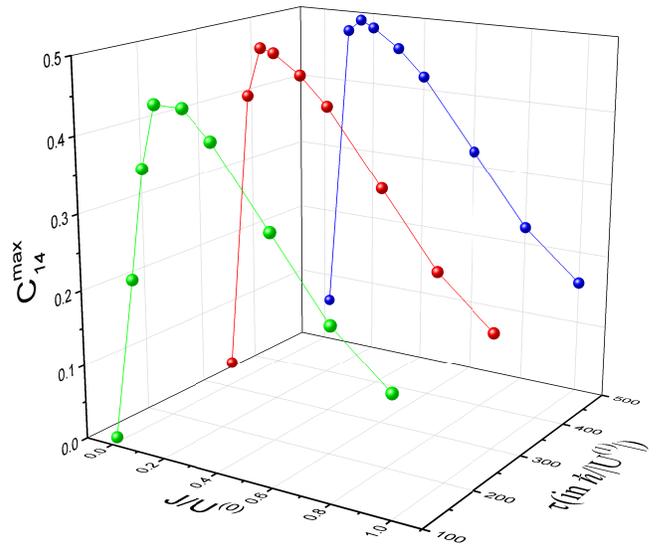}}
\caption{(Color online) Entanglement generation in terms of
cross-correlation as a function of ramp time and initial $J/U^{(0)}$ value
for non-linear ramp of exponent $\alpha=0.8$ and two site
addressing only.} \label{fig7}
\end{figure}


\begin{figure}[hhh]
\rotatebox{0}{\includegraphics*[width=\linewidth]{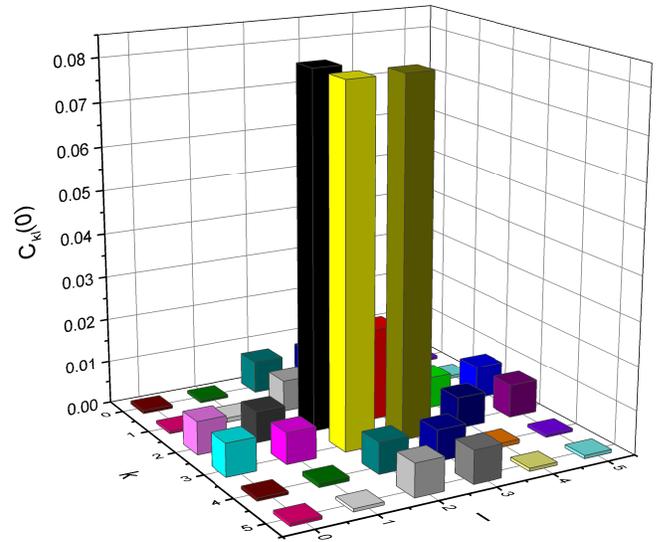}}
\caption{(Color online) The real part of the density matrix for six site linear chain. In this case the correlation
$C_{kl}(0)$ is for $J/U^{(0)}=0.15$ applied to only sites $1$ and $4$ which has been considered as the initial state.} \label{fig8}
\end{figure}

In attempting such generalization, we find that in analog to the
dynamic protocol for cooling of ions, the entanglement generation
protocol also yields the shortest time for $\alpha=0.8$. So in what
follows, we concentrate on a non-linear ramp with $\alpha=0.8$.
Furthermore, in analogy to the cooling protocol, we resort to
two-site addressing, {\it i.e.}, we keep a finite initial
interaction only the on two sites where we change the interaction
parameter; for the rest of the sites, $U_i=0$ at all times. We note
that, as shown in Fig.\ \ref{fig6}, the two-site addressing do not
change the dynamics of $C_{14}(t)$ in any appreciable manner; in
fact, such a protocol leads to faster plateauing of $C_{14}(t)$
which signifies a faster entanglement generation. Thus we shall
restrict ourselves to the two-site addressing protocol for the rest
of this section.

\begin{figure}[hhh]
\rotatebox{0}{\includegraphics*[width=\linewidth]{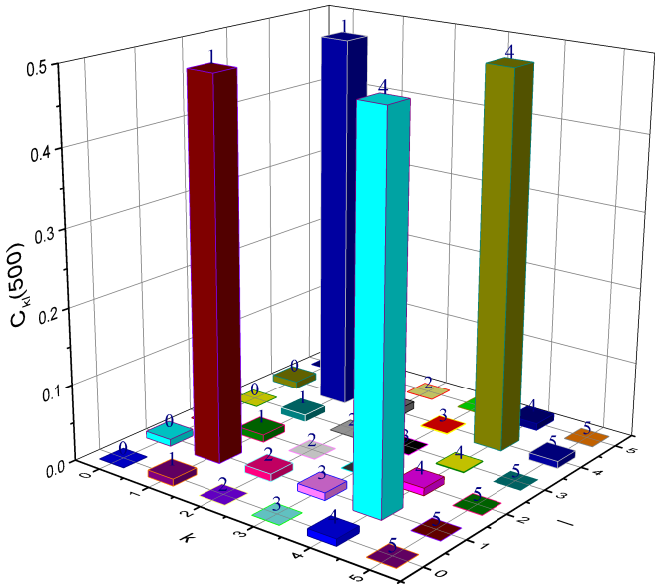}}
\caption{(Color online) Tomograph for $J/|U^{(1)}|=0.15$ with
$U^{(1)}<0$ for $k=1$ and $4$ and $U^{(1)}>0$ otherwise. The other
parameters are $\tau |U^{(1)}|/\hbar=500$ and $\alpha=0.8$.}
\label{fig9}
\end{figure}

Next, we study the maximum attained value of the correlation,
$C_{14}^{\rm max}$, in a given ramp with exponent $\alpha=0.8$ as a
function of the ramp time $\tau$ and $J/U^{(0)}$. The corresponding
plot, shown in Fig.\ \ref{fig7}, indicates that it is indeed
possible to reach very close to the maximal value $1/2$ of $C_{14}$
within a finite time; the optimal parameters for this turns out to
be $\tau= 500\hbar/|U^{(1)}|$ and $J/U^{(0)}=0.15$. To show that such a
correlation indeed represent realization of a Bell state, we compute
the cross correlation between all pairs of sites, $C_{kl}$ for
$\tau= 500 \hbar/|U^{(1)}|$, $\alpha=0.8$ and $J/U^{(0)}=0.15$ for sites $i=1$
and $i=4$. At $t=0$, before the ramp is carried out, we find from
Fig.\ \ref{fig8}, that $C_{kl}(0)$ is finite and has comparable
magnitude for most pair of sites. This situation is to be contrasted
with the behavior of $C_{kl}(t=\tau)$; as shown in Fig.\ \ref{fig9},
$C_{kl}(t=\tau)$ is close of to $1/2$ for $k=1,4$ and $l=1,4$ and is
close to zero otherwise. This is clearly a signature of generation
of Bell state through dynamics. We note that this method may also be
used to generate entangled states involving multiple sites such as
the Greenberger--Horne--Zeilinger state \cite{ks2}; however generation and maintaining
stability of these states are expected to be much more complicated
experimentally.

\begin{figure}[bbb]
\rotatebox{0}{\includegraphics*[width=\linewidth]{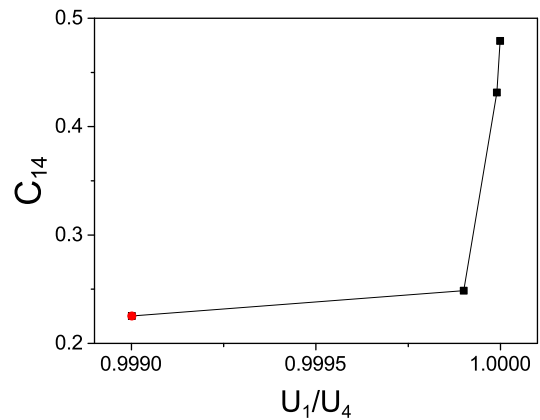}}
\caption{(Color online) Plot of $C_{14}(t_f)$ with the ratio of two
on-site interaction strengths $U_1$ and $U_4$ reflecting the
sensitive dependence of the entanglement on the on-site energy.}
\label{fig10}
\end{figure}

Before ending this section, we would like to make a comment
regarding the requirement that the sites between which such cross
correlation is generated need to be symmetrically placed for a
finite chain. This requirement stems form the fact that the on-site
energies of these two sites need to be identical for generating a
Bell-like state. These on-site energies have contribution from
virtual boson hopping processes which are ${\rm O}(J^2/U^2)$ or
higher. Since the boson hopping processes are cut off by chain ends,
these contributions vary with the position of a site; thus two sites
which are asymmetrically placed with respect to the chain ends would
have different on-site energies due to different contributions from
virtual boson hopping processes. Since for creating entangled states
involving bosons in two different sites, one needs to have equal
on-site energies of the two sites to a high-degree of accuracy,
these sites need to be placed symmetrically with respect to the chain
ends. Indeed, as shown in Fig.\ \ref{fig10}, a variation of on-site
energy induced by changing the ratio $U_1^{(1)}/U_4^{(1)}$ from one, leads to a decay
of the entanglement. This inherent fragile structure of the
entangled state motivates us to at least qualitatively discuss the
effect of the presence of noise (which is an integral part of any
experiment carried in ion trap systems) on entanglement generation.
We shall discuss this in the next section.

\section{Discussion}
\label{dis1}

In this work, we consider cooling and entanglement generation for an
ion trap system with a chain of $L$ sites with $N$ phonons. The
numbers $L$ and $N$ can be controlled and they vary within the range
$2 \le L \le 14$ and $ 0\le N\le 10$ in typical
experimental setups. For cooling, we choose any one ($i^{\rm th}$)
the $L$ sites with $U_i^{(0)}=U^{(0)}$ at $t=0$ and vary $U_i(t)$ according to a
ramp protocol so that $U_i^{(1)}=-U^{(0)}$ at the end of the ramp. This leads
to migration of phonons to the $i^{\rm th}$ sites leading to cooling
of the rest of the sites in the system. We find that $N_i/N \sim
0.97$ for $\tau \simeq 19$ms for a typical $|U^{(1)}|=2.8$kHz and $J/U^{(0)}
\simeq 0.15$. Such a cooling time pertains to a chain of $L=8$; one
expects the cooling time to increase linearly with the chain length
since the migration time of the phonons $\sim L\hbar/J$ increases
linearly with this length for a fixed $J$.

Next, we come to the issue of entanglement generation. As we noted
in Sec.\ \ref{entang1}, the dynamical generation of entanglement
crucially depends on the equality of the on-site energies of the two
chosen sites $i$ and $j$ of the chain; $C_{ij}$ falls down rapidly
from its maximal value of $0.97 C_{ij}^{\rm max}$ when these on-site
interactions differ from each other. Thus the entangled state
generated during dynamics is expected to be fragile against any
relative fluctuation of the on-site interactions $U_i$ and $U_j$.
Such fluctuations are integral part of a realistic experimental
setup since they originate from the fluctuation of the intensity and/or frequency of
the lasers used to generate the on-site interactions. Thus any realistic protocol for
dynamic generation of entangled state needs to address the effect of
noise on $C_{ij}$; in what follows, we qualitatively address this
issue.

\begin{figure}[bbb]
\rotatebox{0}{\includegraphics*[width=\linewidth]{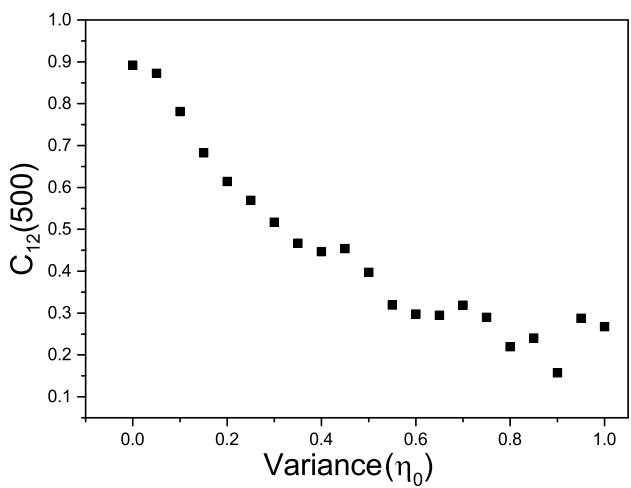}}
\caption{(Color online) Plot of $C_{14}(t_f=500 \hbar/|U^{(1)}|)$ with the
effective noise temperature $\eta_0$ in units of $U_0$ showing the
gradual decay of entanglement with larger noise temperature. The
initial value of $J/U_0$ is chosen to be $0.2$ for this plot.}
\label{fig11}
\end{figure}

Assuming that slow drifts can be controlled by suitable locking, the time scale of the laser fluctuations $\sim 100$ns are much
shorter than typical entanglement generation time $\tau \sim 20$ms.
Thus in any typical measurement, one expects the effect of noise to
self-average; an estimate of $C_{ij}$ can thus be obtained from its
average value over several noise realization. A direct estimate of
$C_{ij}$ thus requires solution of stochastic differential equations
given by Eq.\ \ref{sch3} with $U_i(t) \to U_i(t) + \delta U_i(t)$,
where $\delta U_i(t)$ denotes the random fluctuation of the on-site
interaction on the $i^{\rm th}$ site. We do not attempt to solve
this problem here. Instead, to obtain a qualitative understanding of
the effect of noise, we consider a two site problem with two phonons
($L=2$, $N=2$) which are described by the Hamiltonian
\begin{eqnarray}
H_1(t) &=& J (b_1^{\dagger} b_2 + {\rm h.c}) + U(t) [ {\hat n}_1
({\hat n}_1-1) \nonumber\\
&& + (1+ \delta U/U(t)) {\hat n}_2 ({\hat n}_2-1) ]. \label{twoham}
\end{eqnarray}
Here $b_{1}(b_2)$ are the phonon annihilation operators on sites
$1(2)$, ${\hat n}_i = b_i^{\dagger} b_i$ is the phonon number
operator, $J$ is the hopping amplitude between the sites, $U(t)=
U^{(0)}(1-2(t/\tau)^{\alpha})$ is the on-site interaction which is
varied from $U^{(0)}$ to $-U^{(0)}$ within a ramp time $\tau$, and $\delta
U$ is a random number representing the relative fluctuation between
the on-site interactions at site $1$ and $2$. In what follows we
shall choose $\delta U$ to have a Gaussian distribution
$\exp[-x^2/\eta_0^2]$, where the variance $\eta_0 = k_B T_{\rm eff}$
denotes the effective noise temperature of the system.

The Hilbert space of the above model can be charted out in terms of
the number states of bosons on each site. For simplicity we are assuming that the ions start in the entangled state and the noise is imposed at the two sites of entanglement. For $N=2$, the Hilbert
space consists of three states which are given by $ |0\rangle =
|n_1=1,n_2=1\rangle$, $|1\rangle \equiv |n_1=2, n_2=0\rangle$, and
$|2\rangle = |n_1=0,n_2=2\rangle$. Thus the state for the bosons at
any time $t$ during the dynamics and for a given realization of the
noise $\delta U$ can be written as
\begin{eqnarray}
|\psi(t)\rangle = \sum_{\alpha=0}^2 c_{\alpha}(t) |\alpha\rangle
\label{nstate}
\end{eqnarray}
where the equations for the coefficients $c_{\alpha}(t)$ can be
obtained from the Schrodinger equation $ i \hbar \partial_t
|\psi(t)\rangle = H_1(t) |\psi(t)\rangle$ and are given by
\begin{eqnarray}
i \hbar \partial_t c_0 &=& \sqrt{2}J (c_1 + c_2) \nonumber\\
i \hbar \partial_t c_1 &=& U(t) c_1 + \sqrt{2}J c_0 \nonumber\\
i \hbar \partial_t c_2 &=& [U(t) + \delta U] c_2 + \sqrt{2} J c_0.
\label{nevoleq}
\end{eqnarray}
Eqs.\ \ref{nevoleq} can be solved numerically to obtain
$c_{\alpha}(t)$ for a given disorder realization. The initial
conditions for solving such equations can be obtained from finding
their value for the system ground state at a fixed value of $J/U^{(0)}$.
This can be done by numerical minimization of the system energy $E=
\langle \psi(0)|H(0)|\psi(0)\rangle$.

Having obtained $c_{\alpha}(t)$ for a given disorder realization,
one can compute the cross correlation $C_{12}^d$ given by
\begin{eqnarray}
C_{12}^d (t) &=& \langle \psi(t)|(b_2^{\dagger} b_1)^2
|\psi(t)\rangle/2  = c_1^{\ast} c_2  \label{crosscorr}
\end{eqnarray}
One then averages $C_{12}$ over several disorder realization and
obtain $ C_{12}(t) = 2 {\rm Re} [\langle C_{12}^d(t) \rangle_d]$,
where the average is taken over $M=100$ realization of disorder. The
behavior of $C_{12}(t)$ as a function of the effective noise
temperature $\eta_0$ is shown in Fig. \ \ref{fig11} for
$J_i/U^{(0)}=0.2$ and $t=500 \hbar /|U^{(1)}|$. We note that the
entanglement gradually decays from its value $0.9$ for $\eta_0=0$ to
around $0.2$ for $\eta_0=U^{(0)}$ with increasing $\eta_0$. In a typical
experimental setup, the variation of the laser intensity and/or phase is
around $1\%$ and this leads to around $3\%$ fluctuation in the value
of $U$. This suggest that the experimental noise might lead to a
reduction of $\sim 10\%$ value of the cross correlation as obtained from the simple naive model. A more
detailed analysis of this phenomenon which is expected to provide an
accurate estimate of the entanglement reduction in a more realistic
situation is left for future work.

In conclusion, we have studied the non-equilibrium dynamics of
phonons in an ion trap and have shown that such dynamics, for
carefully chosen protocol, may lead to both cooling and entanglement
generation. We have provided a detailed analysis of theoretical
method used to study such dynamics. Using this, we have identified
the non-linear ramp protocol with exponent $\alpha=0.8$ to be
optimal for both cooling and entanglement generation. Finally we
have provided an estimate of cooling and entanglement generation
times based on our analysis which may be confirmed in experiments
and have also qualitatively discussed the effect of the presence of
laser intensity fluctuations on the entanglement generation.

\section{Acknowledgement}

MM and TD would like to acknowledge the financial support from National Research Foundation-Prime Minister's Office, Singapore and the Ministry of Education, Singapore.

\end{document}